\documentclass[aps,prl,twocolumn,groupedaddress,floatfix,showpacs,showkeys]{revtex4}

\usepackage[dvips]{graphicx}

\begin{document}

\title{Scaling of the turbulence transition threshold in a pipe}

\author{B. Hof}
\author{A. Juel}
\author{T. Mullin}
\affiliation{Manchester Center for Nonlinear Dynamics,\\
University of Manchester, Oxford Road, Manchester, M13 9PL, UK}

\date{\today}

\begin{abstract}
We report the results of an experimental investigation of the
transition to turbulence in a pipe over approximately an order of
magnitude range in $Re$. A novel scaling law is uncovered using a
systematic experimental procedure which permits contact to be made
with modern theoretical thinking. The principal result we uncover
is a scaling law which indicates that the amplitude of
perturbation required to cause transition scales as $O(Re^{-1})$.
\end{abstract}

\pacs{47.20.-k,47.27.-i,47.60.+i}

\keywords{Pipe flow, stability, transition, turbulence.}

\maketitle


The puzzle of why the flow of a fluid  along a pipe is typically
observed to change from laminar to turbulent as the flow rate is
increased has been the outstanding challenge of hydrodynamic
stability for more than a century. The issue is both of deep
scientific and engineering interest since most pipe flows are
turbulent in practice even at modest flow rates. All theoretical
work indicates that the flow is linearly stable \cite{drazin}, i.e
infinitessimal disturbances added to the field will decay and
disappear as they travel along the pipe and the flow will remain
laminar.  It is natural to assume that finite amplitude
perturbations are therefore responsible for triggering turbulence
and these become more important as the non-dimensionalized flow
rate, the Reynolds number, $Re$, increases \cite{kelvin}. A
question which may be asked is, if $\epsilon = \epsilon(Re)$
denotes the minimal amplitude of all finite perturbations that can
trigger transition, and if $\epsilon$ scales with $Re$ according
to
$$
\epsilon = O(Re^{\kern 1pt\gamma}) \eqno (1)
$$
as $Re\to \infty$, then what is the exponent $\gamma$
\cite{trefethen}?\ \ A negative value of $\gamma$ is anticipated
and one substantially less than zero would indicate that the
sensitivity of the laminar flow increases rapidly with $Re$, i.e.
the basin of attraction of the laminar fixed point diminishes
rapidly as $Re$ increases. Current estimates for $\gamma$ suggest
that for shear flows it lies within the range $-1 \geq \gamma \geq
-7/4$ from various model studies \cite{eckhardt,meseguer} and
numerical simulations \cite{shan} where and exponent strictly less
than $-1$ is required for transient growth\cite {waleffe1}. A
significant challenge is to relate this theoretical concept to
observation in a quantitative manner although some limited data is
available \cite{draad}. In this Letter we provide evidence from a
novel experiment which suggests a way forward and provides
striking evidence for an exponent of $-1$ which points to a
generic transition \cite{waleffe1}.

The stability of Hagen--Poiseuille flow in a long circular pipe
has intrigued scientists for more than a century since Reynolds'
\cite{reynolds} experimental investigations. Reynolds showed that
when the parameter we now call the Reynolds number $Re$ was
greater than approximately 2,000 then turbulent flow became the
norm in practice. $Re$ is usually defined as $Re = Ud/\nu$ where
$U$ is the mean speed of the flow, $d$ is the diameter of the pipe
and $\nu$ is the kinematic viscosity of the fluid. Importantly, he
also showed that  if the inlet disturbances to the pipe are
minimized then laminar flow can be maintained to higher flow rates
than if they are not. This finding has been extended in modern
times  to Reynolds numbers of $\approx 100,000$ \cite{pfenniger}
in transient flows by taking extraordinary care.

The process whereby turbulence arises is not understood either in
outline or in detail and this problem is an unresolved scientific
challenge. Any advance towards an understanding  of the
fundamentals involved will have widespread impact on flows of
practical interest. For example, the flows in oil and gas
pipelines are often run inefficiently turbulent to avoid the large
pressure fluctuations of the transitional regime. Moreover, the
control of turbulence is a dream of many practitioners, just as an
understanding of turbulence is the desire of many scientists.

In general terms, pipe flow may be considered as a nonlinear
dynamical system $d{\bf u}/dt = f({\bf u},Re)$ which represent the
Navier Stokes equations subject to appropriate forcing and
boundary conditions. The single parameter $Re$, determines the
dynamical state of the system such that there is one linearly
stable fixed point for all $Re$ and another possible attractor,
turbulence, when $Re > Re_{c}$. Hence when $Re < Re_{c}$ all
initial conditions are attracted to the laminar state which is the
global attractor for the system. When $Re >> Re_{c}$ nearly all
initial conditions give rise to turbulence so that the laminar
state is now a local attractor. In practice, $Re_{c} \approx 2000$
so that all disturbances will decay as $t \rightarrow \infty$ for
values of $Re$ smaller than this.
\begin{figure*}[ht]
\begin{center}

\includegraphics[width=5cm,angle=-90]{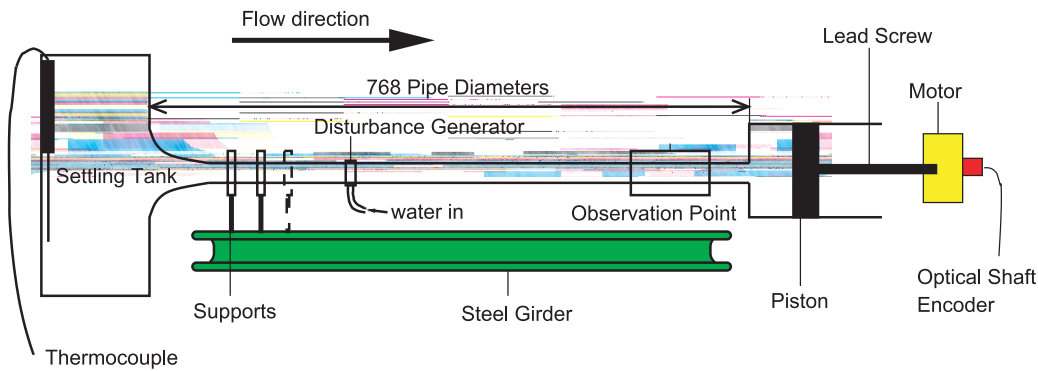}
            \caption{Schematic of the `long pipe' experimental system.}
            \label{exp}
\end{center}
        \end{figure*}
These arguments are consistent both with Reynolds` original
observations and modern experimental results
\cite{wygnanski,draad,han,darbyshire}. Almost all experimental
studies of the problem have been concerned with pressure gradient
driven flows so that large fluctuations in flow rate and hence
$Re$ can, in principle, occur upon transition. In one exception to
this \cite{darbyshire}  a constant mass flux system is used where
the flow is pulled  by a piston and this accurately fixes $Re$.
Impulsive perturbations are used to produce a finite amplitude
stability curve such that disturbances with amplitudes greater
than a threshold produce turbulence while smaller ones decay
downstream. The results are consistent with those obtained with
pressure driven systems \cite{wygnanski,draad,han} so that
localized `puffs' and `slugs' are found at low $Re$ and fully
developed turbulence at larger flow rates.

Modern theoretical research may be broadly split into two
approaches. In one, initially small disturbances on the laminar
state grow in a transient phase \cite{trefethen,grossman} until
they reach a sufficiently large amplitude that nonlinear effects
become important. These ideas have been explored for various
low--dimensional models \cite{bagget} and applied to plane
Poiseuille flow \cite{chapman} and scaling laws for the amplitude
of the perturbation as a function of $Re$ have been provided. An
alternative point of view \cite{waleffe2} is that the turbulent
state originates from instabilities of a finite amplitude solution
which is disconnected from the base state. The basin of attraction
of the turbulent state grows with $Re$ so that any small
perturbation will kick the laminar solution towards it. Such
solutions of the Navier Stokes equations are known to exist in
other flows \cite{barnes,dauchot,anson} but their existence has
not yet been shown in pipe flows.

In drawing a connection between experimental observations and
predictions from models the concept of a perturbation needs to be
defined. In models, the temporal and spatial form of the
perturbation can be accurately specified. On the other hand,
experimentalists rely on injecting and/or subtracting fluid
through slits or holes in an attempt to mimic the mathematical
process. The perturbation can be either periodic \cite{draad,han}
or impulsive \cite{darbyshire} but specifying an essential measure
such as the scale of the amplitude is difficult. Indeed, resolving
the pertinent part of the physical perturbation which gives rise
to transition is in itself a difficult exercise although progress
is being made in that direction \cite{han}. In an attempt to
address this issue  we have devised a novel type of perturbation
which permits a scaling analysis and thereby allows a closer
connection to be made with theory.

 The experimental system can  be regarded as a large computer
controlled syringe which has the capability of pulling water at a
fixed mass flux along a precision bore tube. We have two such
experimental facilities in our laboratory and we will only outline
details of the new rig  as the first system has been described
previously\cite{darbyshire}. The pipe consisted of a 20 mm
diameter Perspex tube which was 15.7 m in length and constructed
from 105 machined sections each of which was 150mm long. The
sections were fitted together and aligned with a laser on a steel
base. Still water was drawn from a tank through a trumpet shaped
inlet into the tube and a similar expansion connected the 260 mm
diameter piston to the tube. The flow state was monitored using
flow visualization and recorded at various spatial locations using
video cameras whose images were stored for further processing.
Laminar flow was achieved for $Re \le 24,000$ verifying the
quality of the construction. We will only report results for $Re
\le 18,000$ and the flow was fully developed at the disturbance
injection point up to $Re=16,000$. A schematic diagram showing the
experimental arrangement for the `long pipe' is given in
Fig.\,\ref{exp} and the shorter pipe was constructed in a very
similar way but was 150 diameters long.

\begin{figure}[ht]
            \begin{center}
            \includegraphics[width=6cm,angle=-90]{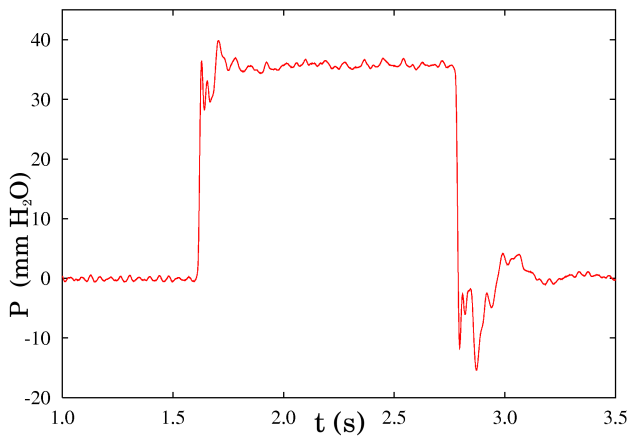}
            \caption{A typical pressure time series for a perturbation of
            amplitude $\Delta p=37\pm1$ mm $H_{2}O$ and width $\Delta t= 1.2\pm 0.01$s.}
            \label{per}
            \end{center}
        \end{figure}

The stability of the flow was probed using a perturbation which
was applied a sufficient number of diameters from the inlet to
ensure fully developed flow over the $Re$ ranges investigated in
the two systems. This distance was 75 and 530 pipe diameters for
the short and long pipes respectively. A single boxcar pulse of
fluid was injected tangentially into the flow via a ring of six
equally spaced 0.5 mm holes to provide the disturbance. The
injection system contained two high speed solenoid valves with
switching times of $\approx 1m\rm{s}$ and the rise and fall times
of the perturbation were limited by the inertia of the piston. The
quantities of fluid injected were in the range 0.01 to 0.1 \% of
the total volume flux where the larger values were required to
cause transition at smaller $Re$.  This novel injection system
enabled us to vary the duration and amplitude of the perturbation
independently. We show in Fig.\,\ref{per} a typical pressure trace
of a perturbation of magnitude ${ \Delta p}$ and width $\Delta t$.
It can be seen that a reasonable approximation to a box-car
function is achieved with relatively small amplitude ringing at
the switching times. The pressure was measured at a single
location to illustrate the form of the perturbation and
measurement of pressure gradient in this format remains a
technical challenge. Hence we use the displaced volume flux
$\Phi_{inj}$ from the injector in our definition of the amplitude
of the perturbation since this sets the spatial extent of the
disturbed flow and defining the amplitude in terms of relative
volume fluxes enables a direct connection with theory. In
principle, the perturbation will have a global effect on the flow
field but checks using injection and suction \cite{darbyshire}
show that it is localized in practice.

\begin{figure}[ht]
            \begin{center}
           \includegraphics[width=6cm,angle=-90]{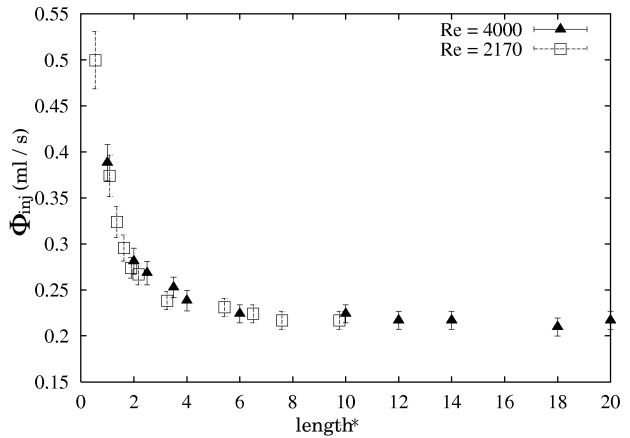}
            \caption{Stability curves measured for $Re = 2170$ and $Re = 4000$.
            Each data point was obtained from forty runs of the experiment and the
            error bars correspond to the widths of the determined probability
            distributions \cite{darbyshire}. The abscissa is the 'length$^*$' of the
            initially disturbed flow in pipe diameters.  }
            \label{stab1}
            \end{center}
        \end{figure}

As discussed above, this means of injecting a disturbance permits
the amplitude and duration of the perturbation to be varied
independently. In Fig.\,\ref{stab1} we show stability curves for
two different values of $Re$ where the amplitude of perturbation
required to cause transition is plotted as a function of the
length of the perturbed flow in pipe diameters. This scale was set
by injecting for a prescribed time and, since the disturbed flow
was advected  at the mean speed of the pipe flow, an estimate of
the spatial extent of the perturbation immediately downstream of
the injection point was made. This is denoted by length$^*$ in
Fig.\,\ref{stab1} and use of this scaling  collapses the two sets
of data onto a single curve.

Perturbations with amplitudes such that they were below the curve
did not cause transition and hence decayed as they propagated
downstream. On the other hand, disturbances which had amplitudes
above the curve  gave rise to sustained disordered flow downstream
which had the form of a localized  `puff' \cite{wygnanski} at
$Re=2170$ or a patch of turbulence at $Re=4000$. As discussed
elsewhere \cite{darbyshire}, the threshold is probabilistic so
that a mean value can be estimated with a narrow  well-defined
width. These are denoted by the error bars in Fig.\,\ref{stab1}
which indicate the width of the experimentally determined
probability distribution of the transition  obtained from forty
rehearsals of the experiment.

The data in Fig.\,\ref{stab1} also shows that the amplitude of
perturbation required for transition is independent of its length
when more than six pipe diameters are initially disturbed. Shorter
length perturbations result in a nonlinear response and these
results are in accord with previous work \cite{darbyshire} where a
short triangular form perturbation was used.

\begin{figure}[ht]
            \begin{center}
           \includegraphics[width=6cm,angle=-90]{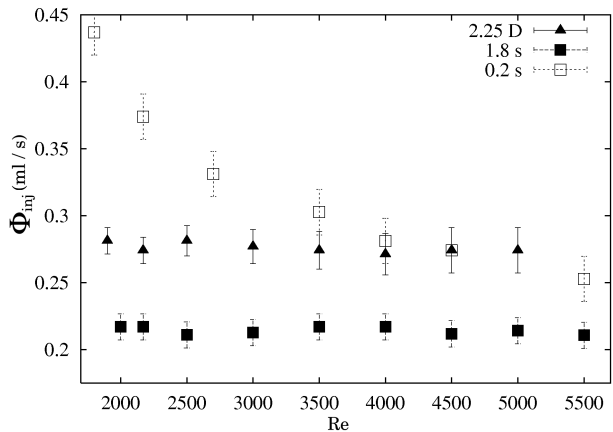}
             \caption{Stability curves measured in the `short' pipe in the range $2,000 \le Re
\le 5,500 $. Fixed duration perturbations were used to obtain the
loci labelled $0.2$ s. and $1.8 $ s respectively. The data set
labelled $2.25 D$ were measured using a variable duration pulse
such that 2.25 diameters of the pipe flow were perturbed.}
            \label{3stab}
            \end{center}
        \end{figure}
We reinforce the above scaling argument with the results shown in
Fig.\,\ref{3stab} which contain data in the range  $2,000 \le Re
\le 5,500 $ where $95\%$ fully developed flow was  achieved with
the shorter pipe. Three stability curves are presented and these
will now be discussed in turn. In the first, a short, 0.2s.,
duration pulse was used and this corresponds to a disturbance
length of one pipe diameter at $Re= 2,000$. This exhibits
nonlinear behavior such that  a rapidly increasing perturbation
amplitude is required to cause transition as $Re = 2,000$ is
approached. The data are in agreement with those obtained in
previous experiments \cite{darbyshire}. In the next, a long
duration (1.8s.) pulse was used and this corresponds to perturbing
nine pipe diameters at $Re=2,000$. Here, the independence of the
amplitude of perturbation required to cause transition discussed
in connection with Fig.\,\ref{stab1} for lengths $\ge 6D$ is
confirmed. In this case, independence  has been uncovered over a
range of $Re$. The final data set was obtained  by varying
duration of the pulse in proportion to the mean flow such that the
length of the flow field which was initially disturbed was kept
constant at $2.25$ pipe diameters. This corresponds to a
perturbation of width 0.45s. at $Re= 2,0 00$. The data reinforces
that scaling of the perturbation by the mean flow is valid. An
alternative way of viewing this data set is to consider it as a
sampled set from a horizontal cut taken through a family of
parallel stability curves, one for each value of the $Re$ set in
the experiment.

 \begin{figure}[ht]
            \begin{center}
           \includegraphics[width=6cm,angle=-90]{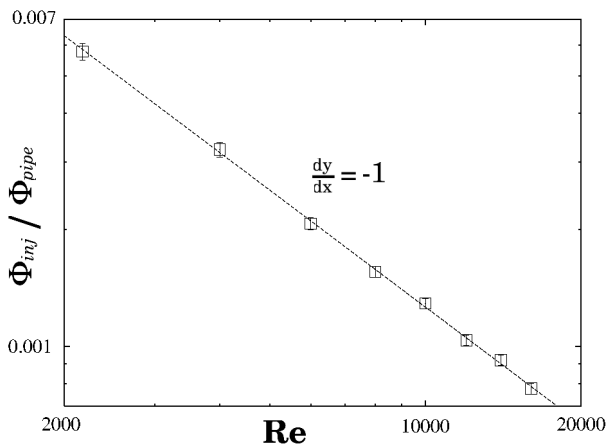}
             \caption{A log-log plot of the stability curve obtained using
             the long pipe. The range of $Re$ covered is 2000 to 18,000 and the
             amplitude of the perturbation has been non-dimensionalized by
             the respective mass flux in the pipe. The least squares fitted line
             has a slope of -1 as indicated. }
            \label{bigp}
            \end{center}
        \end{figure}
It is clear that the two level thresholds cannot continue much
below $Re =2,000$ since experimental evidence suggests that
turbulent flow cannot be maintained below this value
\cite{wygnanski,darbyshire}. It is equally unlikely that the locus
will simply come to an end in parameter space. In this region, we
observe the transient growth of puffs which can persist for many
tens of pipe diameters. This interesting behavior will take
considerable experimental effort to resolve and is the subject of
an ongoing investigation.

An appropriate scaling of the amplitude of the perturbation is the
relative mass flux of the perturbation to that in the pipe.
Clearly, doing this for the two horizontal loci in
Fig.\,\ref{3stab} will produce a proportionality of the form
$O(Re^{-1})$. We next present results from the long pipe in
Fig.\,\ref{bigp} where we were able to test this finding over an
order of magnitude range of $Re$. Here we used a perturbation of
$1.8s$ duration and find the same $O(Re^{-1})$ scaling.

Our investigation  has shown that the amplitude required for
transition in a pipe is constant for a sufficiently long
perturbation. Shorter perturbations show nonlinear dependence but,
these may be scaled with $Re$ so that the  amplitude of
perturbation required for transition  has a $Re^{-1}$ dependence.
This is agreement with recent theoretical estimates for
asymptotically large  $Re$ \cite{jc} which indicates that the
theory may apply over a surprisingly large range of $Re$. Hence we
provide an advance towards forming a closer connection between
modern theoretical approaches and experiment.

\end{document}